\pdfoutput=1
\documentclass[11pt]{article}

\usepackage[preprint]{neurips_2024}
\PassOptionsToPackage{authoryear}{natbib}
\usepackage{longtable}
\usepackage{pifont}
\usepackage{times}
\usepackage{latexsym}
\usepackage{booktabs}
\usepackage{multirow}
\usepackage{graphicx}
\usepackage{enumitem}
\usepackage{algorithm}
\usepackage{algorithmic}
\usepackage{hyperref}       % hyperlinks
\usepackage{url}            % simple URL typesetting
\usepackage{booktabs}       % professional-quality tables
\usepackage{amsfonts}       % blackboard math symbols
\usepackage{nicefrac}       % compact symbols for 1/2, etc.
\usepackage{microtype}      % microtypography
\usepackage{xcolor}    
\usepackage{authblk}

\usepackage{amsmath}
\usepackage[T1]{fontenc}
\usepackage[utf8]{inputenc}

\usepackage{microtype}

\usepackage{inconsolata}

\newcommand{\methodnamews}{\textsc{AgileCoder}}

\newcommand{\methodname}{\methodnamews~}

\title{\raisebox{-0.1cm}{\includegraphics[scale=0.12]{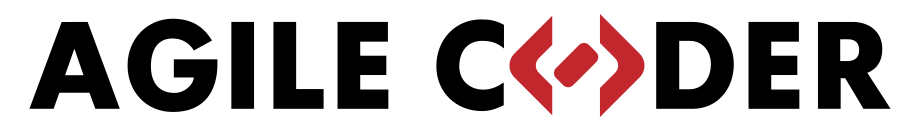}}\ : Dynamic Collaborative 
	Agents \\ for Software Development based on \\ Agile Methodology}

\author{%
	\bf Minh Huynh Nguyen$^{\ast}$, Thang Chau Phan$^{\ast}$, Phong X. Nguyen$^{\ast}$, Nghi D. Q. Bui$^{\ast, \dagger}$ \\
	{$^{\ast}$FPT Software AI Center, Viet Nam} \\
	{$^{\dagger}$Fulbright University, Viet Nam} \\
	\url{https://github.com/FSoft-AI4Code/AgileCoder}
}

\begin{document}
\maketitle
\begin{figure}[!ht]
	\centering
	\makebox[\linewidth]{\includegraphics[width=1.2\linewidth]{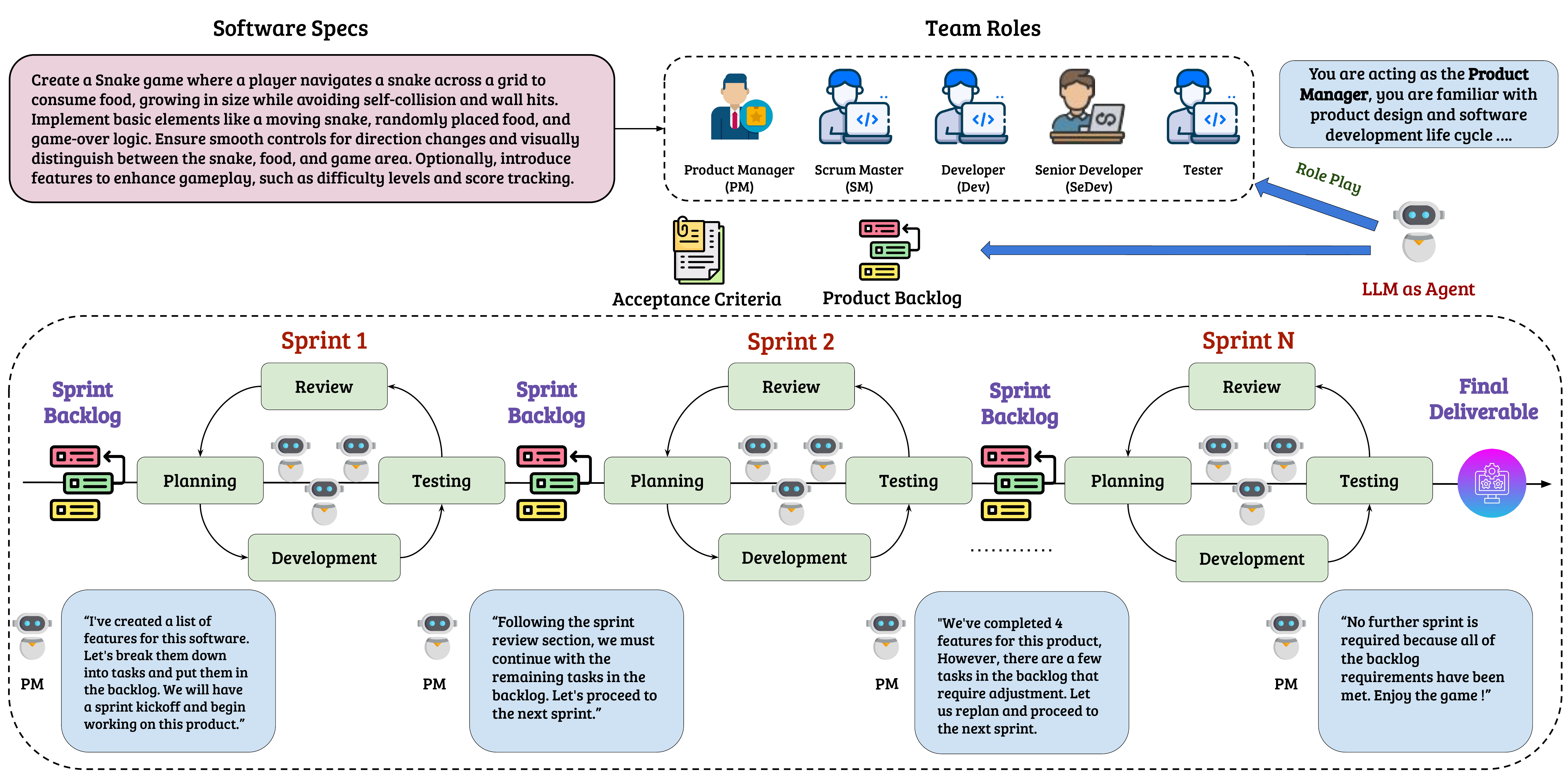}}
	\caption{An overview of \methodnamews}
	\label{fig:overview-agilecoder}
\end{figure}

\begin{abstract}

Software agents have emerged as promising tools for addressing complex software engineering tasks. Existing works, on the other hand, frequently oversimplify software development workflows, despite the fact that such workflows are typically more complex in the real world. Thus, we propose \textbf{\methodnamews}, a multi-agent system that integrates Agile Methodology (AM) into the framework. This system assigns specific AM roles—such as Product Manager, Developer, and Tester—to different agents, who then collaboratively develop software based on user inputs. \methodname enhances development efficiency by organizing work into sprints, focusing on \textit{incrementally} developing software through sprints. Additionally, we introduce Dynamic Code Graph Generator, a module that creates a Code Dependency Graph dynamically as updates are made to the codebase. This allows agents to better comprehend the codebase, leading to more precise code generation and modifications throughout the software development process. \methodname surpasses existing benchmarks, like ChatDev and MetaGPT, establishing a new standard and showcasing the capabilities of multi-agent systems in advanced software engineering environments.

% In the rapidly evolving field of software engineering, achieving efficiency and adaptability in software development processes remains a formidable challenge. Traditional methodologies often fall short in handling the complex, dynamic nature of modern software projects, necessitating innovative approaches that leverage the latest advancements in technology. This paper introduces \methodname, a novel multi-agent software development framework inspired by the principles of Agile methodology. \methodname addresses critical limitations of existing approaches by integrating Large Language Models (LLMs) with a multi-agent system designed to facilitate collaboration, flexibility, and incremental development. Through the Agile Scrum methodology, \methodname overcomes the challenges of conventional software development processes by enabling role-specific agent collaboration, iterative improvements, and adaptability to changing requirements. Our framework demonstrates superior performance on benchmark tasks, establishing a new state-of-the-art and showcasing the potential of multi-agent systems in complex software engineering environments. Our source code is available at {\url{https://github.com/FSoft-AI4Code/AgileCoder}}.
\end{abstract}
\section{Introduction}

Autonomous software agents leveraging Large Language Models (LLMs) offer significant opportunities to enhance and replicate software development workflows~\citep{qian2023chatdev, hong2024metagpt, tang2024collaborative, zhou2023language, huang2023agentcoder}. These agents simulate human software development processes, including design, implementation, testing, and maintenance. MetaGPT~\citep{hong2024metagpt} encodes Standardized Operating Procedures (\textbf{SOPs}) into software development tasks, while ChatDev~\citep{qian2023chatdev} creates a virtual chat-powered technology company; both follow the classic \textbf{waterfall} model. However, these approaches oversimplify the workflow, failing to reflect the dynamic and iterative nature of real-world software development, where approximately 70\% of professional teams adopt \textbf{Agile Methodology (AM)}~\citep{Agile2024}. Moreover, these methods overly depend on LLMs for decision-making and managing code generation, proving inadequate for handling the complexity of entire software repositories, especially when considering repository-level code understanding and generation.
To address these limitations, we propose \methodnamews, a novel multi-agent software development framework based on Agile Methodology (AM). \methodname mimics the AM workflow and adapts it to a multi-agent framework context, allowing for dynamic adaptability and iterative enhancement capabilities. We also introduce a static-analysis-based module called Dynamic Code Graph Generator, which creates a Code Dependency Graph (CDG) that updates whenever the code changes. The CPG serves as a reliable source for agents to retrieve relevant contexts, enabling precise refinements of the software in each sprint.

To evaluate the efficacy of \methodnamews, we conduct an extensive evaluation on two well-known benchmarks, HumanEval \citep{chen2021humaneval} and MBPP \citep{austin2021mbpp} and a software development benchmark, ProjectDev. Experimental results demonstrate that \methodname achieves the best scores of pass@1 across the first two datasets. For instance, using GPT-3.5 Turbo as the backbone model, \methodname achieves 70.53\% and 80.92\% in pass@1 on HumanEval \citep{chen2021humaneval} and MBPP \citep{austin2021mbpp}, respectively, yielding improvements of 7.71\% and 6.19\% compared to the MetaGPT \citep{hong2024metagpt}.  Furthermore, we also collect dataset of software requirements that require the system to produce complex softwares, named ProjectDev. \methodname exhibits outstanding performance on ProjectDev compared to ChatDev \citep{qian2023chatdev} and MetaGPT\citep{hong2024metagpt} in generating executable programs meeting user requirements. 

% In short, our main contributions are as follows:

In short, our main contributions are summarized as follows:

\textbf{(1)} We introduce \methodnamews, a novel multi-agent software development framework inspired by Agile methodology, which emphasizes effective communication and incremental development among agents. This framework allows for the inheritance of outputs across sprints, enabling continuous refinement and increasing the success likelihood of the final products. 

\textbf{(2)} We integrate a static analysis method into the multi-agent workflow through the Dynamic Code Graph Generator (DCGG), which dynamically produces a Code Dependency Graph (CDG). This graph captures the relationships among code components as the codebase evolves, providing a reliable source for agents to retrieve relevant contexts and thereby enhancing the quality of the produced software. Our evaluation demonstrates a significant performance increase in real-world benchmarks when utilizing contexts retrieved from the CDG.

\textbf{(3)} Our evaluations confirm that \methodname achieves new state-of-the-art (SOTA) performance on established benchmarks like HumanEval \citep{chen2021humaneval} and MBPP \citep{austin2021mbpp}, as well as our newly proposed benchmark for real-world software development, named ProjectDev. This framework outperforms recent SOTA models such as MetaGPT \citep{hong2024metagpt} and ChatDev \citep{qian2023chatdev}, demonstrating its efficacy in practical software development scenarios.

% \begin{figure*}[!ht]
% 	\centering
% 	\includegraphics[width=0.85\linewidth]{images/graph_workflow.pdf}
% 	\caption{Illustration of how  Dynamic Code Graph Generator (DCGG) contributes to \methodname during the generation of a Python application.}
% 	\label{fig:graph_workflow}
% \end{figure*}

\section{Related Work}
\paragraph{Deep Learning for Automated Programming}
In recent years, applying deep learning to automated programming has captured significant interest within the research community~\citep{balog2016deepcoder, bui2018hierarchical, bui2021treecaps, feng2020codebert, wang2021codet5, allamanis2018survey, bui2023codetf, guo2020graphcodebert, guo2022unixcoder}. Specifically, Code Large Language Models (CodeLLMs) have emerged as a specialized branch of LLMs, fine-tuned for programming tasks~\citep{wang2021codet5,wang2023codet5+, feng2020codebert, allal2023santacoder, li2023starcoder, lozhkov2024starcoder, guo2024deepseek, pinnaparaju2024stable, zheng2024opencodeinterpreter, roziere2023code, nijkamp2022codegen, luo2023wizardcoder, xu2022systematic, bui2022detect}. These models have become foundational in an industry that offers solutions like Microsoft Copilot, Codium, and TabNine, excelling at solving competitive coding problems from benchmarks such as HumanEval \citep{chen2021humaneval}, MBPP \citep{austin2021program}, and APPs~\citep{hendrycks2021measuring}. Despite achieving good results with benchmark tasks, these models often struggle to generate real-world software that requires complex logic and detailed acceptance criteria, which are essential for practical applications~\citep{hong2024metagpt, qian2023chatdev}.

\paragraph{LLMs-based Multi-Agent Collaboration for Software Development}
Recently, LLM-based autonomous agents have attracted significant interest from both industry and academia~\citep{wang2024survey, guo2024large, du2023improving}. In software development, the deployment of agent-centric systems specialized in coding tasks has led to notable advancements~\citep{hong2024metagpt, qian2023chatdev, chen2023agentverse, huang2023agentcoder, zhong2024ldb, lin2024llm, yang2024swe}. These systems feature distinct roles—Programmer, Reviewer, and Tester—each dedicated to a specific phase of the code generation workflow, thereby enhancing both quality and efficiency. 
Accompanying these agent-centric systems are benchmarks designed to evaluate their ability to handle real-world software engineering tasks. For instance, SWE-Bench~\citep{jimenez2023swe} challenges multi-agent systems to resolve real GitHub issues. Similarly, MetaGPT introduces SoftwareDev~\citep{hong2024metagpt}, a suite of software requirements from diverse domains that require agents to develop complete software solutions. In our case, we also compile a diverse suite of software requirements to benchmark the capability of ~\methodname to produce real-world software effectively.
\section{Background}
\subsection{Agile Methodology for Professional Software Development}

Agile, derived from the Agile Manifesto~\citep{agilemanifesto2001}, is a flexible software development methodology that emphasizes pragmatism in delivering final products. It promotes continuous delivery, customer collaboration, and swift adaptation to changing requirements. Unlike traditional linear methods such as the Waterfall model~\citep{waterfall}, Agile employs iterative development through \textit{\textbf{sprints}}—short cycles that enable rapid adjustments and frequent reassessment of project goals. This iterative approach enhances alignment with customer needs and fosters open communication and shared responsibility within teams. Agile’s adaptability makes it especially effective for managing complex projects where requirements may evolve over time. By integrating Agile principles with collaborative agents in software development, we offer a novel perspective to designing multi-agent systems.

\subsection{Repository-Level Code Understanding \& Generation}
Generating code at the repository level is a significant challenge for Large Language Models (LLMs) in real-world software engineering tasks \citep{shrivastava2023repofusion, shrivastava2023repository, bairi2023codeplan, zhang2024codeagent, agrawal2023guiding, phan2024repohyper}. Real-world codebases are complex, with interconnected modules, and as the context size increases, LLMs face limitations. This has led to research on selecting relevant contexts~\citep{luo2024repoagent, shrivastava2023repofusion, liu2023repobench} and optimizing their use. Software agents, like ChatDev and MetaGPT, aim to generate fully functional, executable software comprising various files, classes, and modules, rather than just solutions for simple tasks like those in HumanEval~\citep{chen2021humaneval} or MBPP~\citep{austin2021program}. This requires agents to understand all existing contexts, including files, classes, functions, and libraries, when generating code or fixing bugs. However, this need for comprehensive repository-level code understanding and generation has often been overlooked in prior research.
\section{\methodnamews: An Agentic Framework for Software Development}
\begin{figure*}[!ht]
	\centering
	\includegraphics[width=\linewidth]{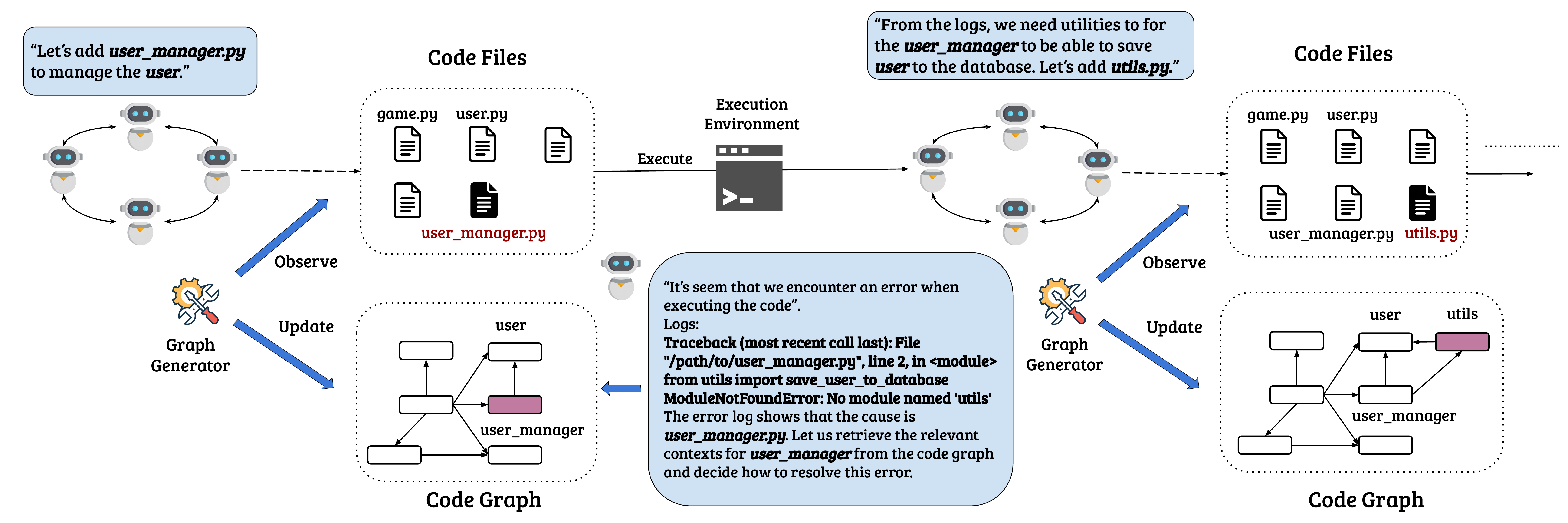}
	\caption{Illustration of how  Dynamic Code Graph Generator (DCGG) contributes to \methodname during the generation of a Python application.}
	\label{fig:graph_workflow}
\end{figure*}

Figure~\ref{fig:overview-agilecoder} presents an overview of \methodnamews, which employs multiple agents in roles such as Product Manager, Scrum Master, Developer, Senior Developer, and Tester, collaborating through an Agile Methodology-inspired workflow. The development process incorporates the Execution Environment for running code during testing and the Dynamic Code Graph Generator (DCGG) (Figure~\ref{fig:graph_workflow}) for dynamically generating the Code Dependency Graph whenever the code is updated. The Execution Environment provides tracebacks to agents for code refinement, while the DCGG enables agents to retrieve relevant contexts for accurate code generation and correction. Detailed descriptions of these modules are provided in the following Sections.

\subsection{Agent Roles}
\label{sec:agent_roles}
The roles of each agent are defined as follows:
\begin{itemize}[leftmargin=*]
	\item Product Manager (PM): Takes requirements from users and creates a product backlog, which includes development tasks and acceptance criteria.
	\item Scrum Master (SM): Provides backlog feedback to the PM to enhance sprint achievability.
	\item Developer (Dev): Primarily focuses on developing tasks, which include generating and refactoring code based on feedback from other agents.
	\item Senior Developer (SD): Reviews code generated by developers, provides feedback for code refactoring, and ensures quality control.
	\item Tester: Generates test cases and scripts to validate the code developed by the developers.
\end{itemize}

\subsection{Overview of Workflow}
\label{sec:workflow}
The workflow starts with the Product Manager (PM) initiating backlog planning after receiving user requirements. The PM writes tasks and acceptance criteria in the backlog. The Scrum Master (SM) reviews the backlog, assesses task feasibility, and may request revisions from the PM. The output is the \textit{product backlog}, capturing all tasks required for final deliverables. Once planning is complete, the SM initiates development with a sprint.
Each sprint includes Planning, Development, Testing, and Review phases. During Planning, the SM selects tasks from the Product Backlog for the Sprint Backlog. The remaining phases involve sub-phases where agent pairs collaborate on tasks (details in Section 4.2). After all phases, the SM evaluates progress to decide if the software is ready for delivery. If not, the next sprint begins with Planning. This repeats until the SM decides the software is deliverable, at which point a termination signal concludes the development pipeline.

\subsubsection{Planning Phase}
At the start of each sprint, the Product Manager (PM) drafts a plan including tasks selected from the product backlog, defined acceptance criteria, and insights from reviews of previous sprints, if applicable. The Scrum Master provides feedback to refine this plan, ensuring that it aligns with the project’s current objectives and constraints. The output of this phase is the \textbf{sprint backlog}, outlining the scope of tasks to be completed within the sprint.

\subsubsection{Development Phase}\label{subsec:implementation}
Following the planning phase, the PM directs the Developer to begin implementation. To enhance clarity for other agents, the Developer is required to annotate each method/function with docstrings. However, due to potential inaccuracies from LLM hallucinations \citep{manakul2023selfcheckgpt}, the code produced might not always align with the sprint backlog and acceptance criteria. To mitigate this, the Senior Developer employs a static review process inspired by peer review practices. This review focuses on identifying bugs, logic errors, and edge cases, subsequently providing feedback for corrections.
The review process is structured into three sequential steps to manage complexity and improve the effectiveness of feedback. The first step involves checking for basic errors like empty methods and missing import statements, followed by ensuring that source code fulfills the sprint backlog with the final step confirming that source code meets acceptance criteria and is free of bugs.
% \begin{enumerate}[leftmargin=*]
% 	\item \textbf{Basic Implementation Checks:} Verification of implementation basics, such as the presence of empty methods (e.g., \textit{pass} in Python), adequacy of docstrings, and completeness of import statements.
% 	\item \textbf{Backlog Compliance:} Assessment of whether source code fulfills the tasks listed in the sprint backlog and includes all specified features.
% 	\item \textbf{Criteria Satisfaction and Bug Identification:} Final review to ensure the code meets the sprint's acceptance criteria and is free from potential bugs.
% \end{enumerate}

By breaking down the code review into these structured steps, it becomes feasible for LLMs to conduct thorough static analyses and provide actionable feedback, thereby enhancing the accuracy and reliability of the development phase. An ablation study can be found in the Appendix \ref{appendix:code-review}.
% Ablation study on this prompting strategy can be found in the Appendix \ref{}.

\subsubsection{Testing Phase}
Despite thorough review, error-free code cannot be guaranteed due to LLM hallucinations. Therefore, a tester is employed to write test suites and implement a testing plan, furnishing real-time feedback to the Developer for iterative code refinement. 

\paragraph{Writing Test Suites}
During a sprint, we inherit source code from previous sprints and implement new features. While existing code undergoes thorough review and testing, new code lacks such scrutiny, necessitating the creation of test cases to ensure its functional correctness.  We utilize the code graph $G$ created by the DCGG (Section~\ref{sec:dcgg}) and the list of changed files $\mathcal{F}$ to find files requiring testing.  This process can be formally described as $\bigcup f(n_i) \cup \{n_i\}, n_i \in \mathcal{F}$, where $f$ returns ancestor nodes in the graph $G$ of an input node. For instance, in Figure \ref{fig:graph_workflow}, if the file \textit{user\_manager.py} undergoes any changes, we should recheck its functional correctness along with that of its ancestor files, but not \textit{user.py}. The Tester is then responsible for writing test suites for all necessary files.

\paragraph{Writing A Testing Plan}
After writing test cases, we have multiple testing scripts that must be executed in a specific order to avoid inconsistency among code files and unnecessary costs. For instance, in Figure \ref{fig:graph_workflow}, the file \textit{user\_manager.py} depends on the file \textit{user.py}, making it illogical to test \textit{user\_manager.py} before \textit{user.py}.  Fortunately, we can obtain a logical testing plan by reversing a \textit{\textbf{topological order}} among testing scripts of the code graph $G$. Furthermore, we want the final software to be executable, so the Tester is required to write commands to evaluate its executability.

\paragraph{Fixing bugs}
% Once obtaining a well-defined testing plan, we run files iteratively in line with the testing plan until a problem like a bug or a failed test case is raised. After that, the Programmer will resolve issues based on test reports provided by the Tester. The process will repeat until the testing plan is finished.

Once a well-defined testing plan is established, files are iteratively executed according to the plan until issues such as bugs or failed test cases arise. The Developer addresses these issues based on test reports provided by the Tester, repeating the process until the testing plan is completed.

% Moreover, the inclusion of complete source code in instructions can be redundant and detrimental to the performance. To mitigate this problem, we utilize tracebacks and the graph G to select relevant files associated with errors. We retrieve file names appearing in tracebacks; however, they sometimes do not mention involved files. This problem can be solved by getting node dependencies of the node corresponding to the last file name appearing in the tracebacks since it is the place where bugs happen and it includes missing relationships.

\subsubsection{Review Phase}
At the end of a sprint, the Tester runs software to write a testing report. The Product Manager then collects the sprint backlog, source code, and the testing report to assess completed, failed, and incomplete tasks. This information accumulates over sprints to form an overall report. The Product Manager then compares this overall report with the product backlog and acceptance criteria to decide whether to conclude the task or plan the next sprint. If concluding, the Scrum Master writes detailed documentation, including how to run and install necessary libraries. If planning another sprint, the Product Manager reviews the product backlog, acceptance criteria, and the current overall report to create the next sprint plan.

\subsection{Dynamic Code Graph Generator for Context-Aware Code Retrieval}\label{subsec:code-graph}
\label{sec:dcgg}
% \begin{figure}
% 	\centering
% 	\includegraphics[width=\linewidth]{images/dependency_graph.png}
% 	\caption{An example of the dependency graph used by the DCGG}
% 	\label{fig:dependency-graph}
% \end{figure}
The Dynamic Code Graph Generator (DCGG) is a critical module in our system that dynamically creates and updates a code dependency graph, denoted as $G$. This graph effectively models the relationships among code files, facilitating efficient context-aware code retrieval. In $G$, each node represents a code file, and each edge indicates a dependency relation, such as the relationship between \textit{user\_manager.py} and \textit{user.py} shown in Figure \ref{fig:graph_workflow}. We primarily capture \textit{import} relationships to maintain the graph's simplicity and efficiency. As the codebase evolves, whether through the addition of new features or corrections of bugs, $G$ is updated to reflect the current state accurately. This is essential for maintaining the integrity of the codebase and ensuring that all changes are properly documented and integrated. Additionally, files modified during a sprint, identified as $\mathcal{F}$, are recorded and linked within $G$ to track all affected dependencies. In summary, the DCGG serves two primary functions:
\begin{enumerate}[leftmargin=*]
	\item \textbf{Assisting in writing test cases and testing plan} When new code has been generated, DDCG ensures that only influenced files need testing. Also, a reasonable testing plan is always obtained by using the $G$.
	\item \textbf{Providing Context for Code Repair:} When an error occurs during code execution, the agent traces back to identify the relevant files using traceback analysis. It retrieves cross-file contexts from the code graph to pinpoint the source of the error, ensuring effective code retrieval.
\end{enumerate}

Existing approaches, such as those employed by ChatDev \citep{qian2023chatdev} and MetaGPT \citep{hong2024metagpt}, often involve loading the entire codebase into the LLMs, which becomes impractical as the codebase grows beyond the token limits of LLMs. DCGG addresses this limitation by maintaining a dependency graph that allows the agent to selectively retrieve relevant information, thereby optimizing the LLM's token usage and maintaining the relevance and accuracy of the information provided to the developers or agents. 

This design approach not only improves the efficiency of code generation and debugging but also ensures that the development process is streamlined and focused, avoiding the overload of irrelevant data and enhancing the overall accuracy of the development workflow. Figure \ref{fig:graph_workflow} illustrates how DCGG updates in response to codebase changes, such as the addition of new classes or relationships, thereby maintaining an up-to-date and accurate reflection of the code structure.

\subsection{Communication Mechanism Among Agents}
Our system employs a dual-agent conversation design across all phases, simplifying the interaction model to just two roles: an instructor and an assistant. This structure avoids the complexities associated with multi-agent topologies and streamlines the consensus process. The instructor is responsible for providing clear instructions and guiding the flow of the conversation toward the completion of specific tasks, while the assistant uses their skills and knowledge to execute the tasks, with interactions continuing until a consensus is reached.

To ensure continuity and coherence in the dialogue between messages, we employ a Message Stream $\mathbf{S}$. This stream acts as a working memory that stores all exchanged messages, allowing agents to create responses that seamlessly connect with the conversation history. Formally, if $i_t$ and $a_t$ are messages produced by the instructor and assistant at time step $t$ respectively, then $\mathbf{S}_t$ is defined as $
\mathbf{S}_t = [(i_1, a_1), (i_2, a_2), ..., (i_t, a_t)]
$.

At the next time step $t+1$, the instructor reviews $\mathbf{S}_t$ to assess alignment of action $a_t$ with provided instructions before issuing further guidance $i_{t+1}$. Simultaneously, the assistant, upon receiving the new instruction $i_{t+1}$, crafts a suitable response $a_{t+1}$. If $a_{t+1}$ satisfies the termination criteria or if the interaction reaches a predefined limit of exchanges, the dialogue concludes. The final message $a_{t+1}$ is stored for future reference, ensuring retention of pertinent and constructive information.

\paragraph{Communication Protocol}
In line with established practices from prior research \citep{qian2023chatdev}, our communication interface utilizes unconstrained natural language. This flexibility allows for easy modifications of prompts, such as adjusting output constraints or changing formats, without significant system-wide impacts. Moreover, \methodname incorporates a technique known as prompt engineering at the onset of each conversation to optimize understanding and task execution. This initial setup ensures that both agents are well-informed about the task requirements and objectives, facilitating the generation of relevant responses aimed at successfully completing the tasks. The agents proceed autonomously without human intervention until consensus is achieved.

\paragraph{Global Message Pool}
To facilitate smooth information flow throughout the system, a global message pool within a shared environment stores the outputs from all conversations. This environment also logs the status of tasks—whether they are completed, pending, or failed—providing a rich context for the agents to make informed decisions. However, given the potential volume of information, direct exposure of all data to the agents could lead to information overload. Therefore, each conversation accesses only the relevant segments of data from the global message pool necessary for task resolution. For instance, while the Product Manager may focus on the product backlog and task statuses for sprint planning, the Developer might need access primarily to specific sections of the code relevant to bug fixes. This targeted access strategy prevents data overload and ensures that all agents operate with the most current and relevant information available.

\section{Experiments}
\subsection{Empirical Results}\label{subsec:results}
\paragraph{Datasets}

\begin{table*}
\centering
\caption{Comparative results on HumanEval and MBPP datasets for various LLMs and LLM-based agents, highlighting performance enhancements achieved through the application of \methodnamews.}
	\label{tab:result_humaneval_mbpp}
\resizebox{\textwidth}{!}{
\begin{tabular}{lllcc} 
\toprule
\multirow{2}{*}{\textbf{Category}} & \multirow{2}{*}{}                    & \multirow{2}{*}{\textbf{Model}} & \multicolumn{2}{c}{\textbf{Dataset Performance}}  \\ 
\cmidrule{4-5}
                                   &                                      &                                 & \textbf{HumanEval} & \textbf{MBPP}                \\ 
\midrule
\multirow{6}{*}{LLMs (prompting)}  & \multirow{6}{*}{}                   
                                                                   & CodeGeeX-13B                    & 18.9               & 26.9                         \\
                                   &                                      & PaLM Coder-540B                 & 43.9               & 32.3                         \\
                                   &                                      & DeepSeeker-33B-Inst             & 79.3               & 70.0                         \\
                                   &                                      & GPT-3.5 Turbo                   & 60.3               & 52.2                         \\
                                   &                                      & Claude 3 Haiku                  & 75.9               & 80.4                         \\
                                   &                                      & GPT 4                           & 80.1               & 80.1                         \\ \midrule
\multirow{7}{*}{LLMs-based Agents} & \multirow{3}{*}{with GPT-3.5 Turbo}  & ChatDev                         & 61.79              & 74.80                        \\
                                   &                                      & MetaGPT                         & 62.80              & 74.73                        \\
                                   
                                   &                                      & \methodname      & \textbf{70.53}     & \textbf{80.92}               \\ \cmidrule{2-5}
                                   & \multirow{2}{*}{with Claude 3 Haiku} & ChatDev                         & 76.83              & 70.96                        \\
                                   &                                      & \methodname      & \textbf{79.27}     & \textbf{84.31}               \\ \cmidrule{2-5}
                                   & \multirow{2}{*}{with GPT 4}          & MetaGPT                         & 85.9               & 87.7                         \\
                                   &                                      & \methodname      & \textbf{90.85}     & -                            \\
\bottomrule
\end{tabular}
}

 \end{table*}

For evaluation, we selected two types of datasets. The first includes well-known benchmarks for assessing code generation capabilities in CodeLLMs: HumanEval \citep{chen2021humaneval} and MBPP \citep{austin2021mbpp}. More experimental settings can be found in the Appendix \ref{subsec:appendix-setting}. These benchmarks primarily feature competitive-level problems that do not fully represent the complexities of real-world software development.

To address this gap, we have collected a collection of 14 representative examples of more intricate software development tasks, collectively referred to as ProjectDev. These tasks cover diverse areas such as mini-games, image processing algorithms, and data visualization. Each task comes with a detailed prompt and requires the system to generate a comprehensive codebase consisting of multiple executable files. We run each task three times and report average numbers. A detailed evaluation process can be found in the Appendix \ref{appendix:evalution-projectdev}.

\paragraph{Metrics}
For HumanEval \citep{chen2021humaneval} and MBPP \citep{austin2021mbpp}, we adopt the unbiased pass@k metric \citep{chen2021humaneval}, following the approach of prior studies \citep{hong2024metagpt, qian2023chatdev}, to evaluate the functional accuracy of top-1 generated code. For the ProjectDev dataset, we focus on practical application and evaluate performance through human assessment and statistical analysis. Human evaluation involves checking the executability of the generated software against expected requirements to determine the success rate in meeting those requirements (e.g. if a generated program is executable and meets 4 out of 10 requirements, its executability rate is 40\%) and computing the total number of errors (\#Errors) when generated programs fail to run. Statistical analysis includes metrics such as runtime, token usage, expenses for all methods, and the average number of sprints (\#Sprints) for only \methodnamews.

\paragraph{Baselines}
We employ SOTA CodeLLMs as baselines, including CodeGen \citep{nijkamp2022codegen}, CodeGeeX \citep{zheng2023codegeex}, PaLM Coder \citep{chowdhery2023palm}, DeepSeek-Coder \citep{guo2024deepseek}, GPT-3.5, GPT-4 \citep{achiam2023gpt4}, and Claude 3 Haiku \citep{anthropic2024introducing}. Given that \methodname is a multi-agent system, we also compare it against leading multi-agent systems used for software development tasks, such as MetaGPT \citep{hong2024metagpt} and ChatDev \citep{qian2023chatdev}. 

\begin{table}[!ht]
    \centering
      \caption{Results on ProjectDev}
    \label{tab:project-dev}
    \begin{tabular}{lccc}
    \toprule
        Statistical Index & ChatDev & MetaGPT & \methodnamews \\
        \midrule
         Executability&32.79 &7.73 &\textbf{57.79} \\ 
         Entire Running Time (s) &120 &\textbf{48} &444\\
       Avg. Time/Sprint (s) &- &- &306\\
         \#Sprints &- &-&1.64\\
         Token Usage &7440 &\textbf{3029}&36818\\
         Expenses (USD) &0.12 &\textbf{0.02} &0.44\\
         \#Errors &6 &32 &\textbf{0}\\
         \bottomrule
    \end{tabular}
  
\end{table}
\paragraph{Results} Table \ref{tab:result_humaneval_mbpp} shows that \methodname significantly outperforms recent SOTA multi-agent frameworks and CodeLLMs on the HumanEval and MBPP benchmarks. \methodname achieves an average improvement of 5.58 and 6.33 in pass@1 over ChatDev and MetaGPT on HumanEval, respectively, with similar improvements on MBPP.
Results on ProjectDev (Table \ref{tab:project-dev}) further demonstrate \methodnamews's superiority in software development tasks. \methodname shows substantial improvements in executability over ChatDev and MetaGPT, without producing any non-executable programs. These advantages can be attributed to \methodnamews's incorporation of the planning phase, generated test cases, and efficient code retrieval, which ChatDev and MetaGPT lack. Although \methodname requires more tokens and running time due to the inherent complexity of the Agile Scrum methodology, it efficiently completes user tasks in an average of 1.64 sprints. Our experiments consistently demonstrate \methodnamews's superiority across various benchmarks.

\textit{It is important to acknowledge that HumanEval and MBPP might not be the most suitable benchmarks for evaluating such complex multi-agent systems, as they primarily contain simple problems for competitive programming—an issue also recognized by previous work~\citep{hong2024metagpt}. As such, benchmarks like ProjectDev are more appropriate for assessing the performance of these systems. We are aware that MetaGPT presents a similar benchmark named SoftwareDev, and ChatDev manually crafts a benchmark called the Software Requirement Description Dataset (SRDD) for the same purpose. Unfortunately, neither of these benchmarks is publicly available. In contrast, our dataset will be released publicly to facilitate open research in this domain. }

% \bottomrule
% \begin{table}[ht]
% \centering
% \begin{tabular}{lcc}
% \toprule
% Dataset               & Method     & Pass@1 \\ \midrule
% \multirow{4}{*}{MBPP} & GPT 3.5 & 52.20\\
% &ChatDev    & 74.80  \\
%                          & MetaGPT    & 74.73  \\
%                          & \methodnamews & \textbf{76.27}  \\
%                          \midrule
%  \multirow{4}{*}{HumanEval} & GPT 3.5 & 60.30\\
%   & ChatDev  & 61.79 \\
%  & MetaGPT    & 62.80  \\
%  & \methodnamews & \underline{64.63}  \textbf{70.51} \\
% % \multirow{2}{*}{GPT 4}   & MetaGPT    &        \\
% %                          & AgileCoder &       
% \bottomrule
% \end{tabular} 
% \caption{Results on MBPP and HumanEval with GPT-3.5 model}
% \label{tab:result_humaneval_mbpp}
% \end{table}

% \begin{table}[ht]
% \centering
% \begin{tabular}{lc}
% \toprule
% Model/Method                & Pass@1 \\ 
% \midrule
% AlphaCode-1.1B & 17.1\\
% Incoder-6.7B & 15.2 \\
% CodeGeeX-13B &18.9 \\
% PaLM Coder-540B & 36.0\\
% WizardCoder-33B-1.1 & 73.8\\
% DeepSeeker-33B-Inst & 78.0\\
% \midrule
% GPT-4 & 82.3\\
% MetaGPT & 85.9\\
% \methodnamews & \textbf{87.4}\\

% % \multirow{2}{*}{GPT 4}   & MetaGPT    &        \\
% %                          & AgileCoder &       
% \bottomrule
% \end{tabular} 
% \caption{Results on HumanEval}
% \label{tab:result_humaneval_gpt4}
% \end{table}

\subsection{Analysis}

\paragraph{Impact of The Number of Sprints} We conduct an ablation study to assess the impact of incremental development.  Incremental development involves multiple sprints, whereas its removal condenses the process into a single sprint. Results in Table \ref{tab:ablation_incremental_dev_code_review} show that incremental development leads to better performance on HumanEval and MBPP across two models, including GPT-3.5 Turbo and Claude 3 Haiku. This advantage stems from inheriting outputs from previous sprints for further refinement and addressing existing issues in subsequent iterations, thereby increasing the likelihood of successful outcomes.

% Additionally, it is noteworthy that in the setting of the lack of incremental development, this variant surpasses ChatDev \citep{qian2023chatdev} and MetaGPT \citep{hong2024metagpt} {\color{red}due to the contribution of planning and code review.}

% \usepackage{booktabs}
% \usepackage{multirow}

\begin{table}[!ht]
\centering
\caption{Ablation study on the incremental development, code review and writing test suite}
\label{tab:ablation_incremental_dev_code_review}
\begin{tabular}{llcc}
\toprule
\multirow{2}{*}{}                   & \multirow{2}{*}{Model} & \multicolumn{2}{c}{Dataset} \\ \cmidrule(l){3-4} 
                                    &                        & HumanEval       & MBPP      \\ \midrule
\multirow{4}{*}{GPT-3.5 Turbo} 
                                    & \methodnamews             &       70.53          &     80.92      \\ 
                                    & w/o     incremental dev.        &      69.51  (-1.02)         &       78.45 (-2.47)   \\ 
                                      & w/o     writing test suite        &    62.20 (-8.33)            &    75.64 (-5.28)       \\
                                    & w/o     code review        &     68.90 (-1.63)          &       75.41 (-5.51)     \\ \midrule
\multirow{4}{*}{Claude 3 Haiku}       
                                    & \methodnamews             &      79.27           &    84.31       \\ 
                                     & w/o     incremental dev.             &       76.83 (-2.44)           &    82.20 (-2.11)       \\
                                     & w/o     writing test suite        &     73.17   (-6.10)         &    79.86 (-4.45)       \\
& w/o     code review        &         75.00 (-4.27)      &    80.56 (-3.75)        \\ 
                                     \bottomrule      
\end{tabular}

\end{table}

% \begin{table}[ht]
% \resizebox{\linewidth}{!}{
% \begin{tabular}{llcc}
% \toprule
% \multirow{2}{*}{}                   & \multirow{2}{*}{Model} & \multicolumn{2}{c}{Dataset} \\ \cmidrule(l){3-4} 
%                                     &                        & HumanEval       & MBPP      \\ \midrule
% \multirow{3}{*}{GPT-3.5 Turbo} 
%                                     & \methodnamews             &       70.51          &     80.92      \\ 
%                                     & w/o     writing test suite        &    62.20 (-8.31)            &    75.64 (-5.28)       \\
%                                     & w/o     code review        &     68.90 (-1.61)          &       75.41 (-5.51)     \\
%                                     \midrule
% \multirow{3}{*}{Claude 3 Haiku} 
% & \methodnamews             &       79.27          &     84.31      \\ 
% & w/o     writing test suite        &     73.17   (-6.10)         &    79.86 (-4.45)       \\
% & w/o     code review        &         75.00 (-4.27)      &    80.56 (-3.75)        \\ 
%                                     \bottomrule      
% \end{tabular}
% }
% \caption{Ablation study on code review and writing test suite}
% \label{tab:ablation_review_writing_tests}
% \end{table}

\paragraph{Impact of Code Review and Writing Testing Suite}The results in Table \ref{tab:ablation_incremental_dev_code_review} demonstrate the importance of code review and writing test cases in \methodnamews. Removing either of these tasks from sprints leads to performance degradation. In particular, the absence of generated test cases significantly impacts performance, confirming their role in detecting potential bugs and improving code quality through the bug-fixing process. Additionally, code review positively contributes to performance, as LLMs may perform static code analysis and identify bugs.
% \begin{table}[ht]
% \resizebox{\linewidth}{!}{
% \begin{tabular}{llcc}
% \toprule
% \multirow{2}{*}{}                   & \multirow{2}{*}{Model} & \multicolumn{2}{c}{Dataset} \\ \cmidrule(l){3-4} 
%                                     &                        & HumanEval       & MBPP      \\ \midrule
% \multirow{3}{*}{GPT-3.5 Turbo} 
%                                     & \methodnamews             &       70.51          &     80.92      \\ 
%                                     & w/o     writing test suite        &    62.20 (-8.31)            &    75.64 (-5.28)       \\
%                                     & w/o     code review        &     68.90 (-1.61)          &       75.41 (-5.51)     \\
%                                     \midrule
% \multirow{3}{*}{Claude 3 Haiku} 
% & \methodnamews             &       79.27          &     84.31      \\ 
% & w/o     writing test suite        &     73.17   (-6.10)         &    79.86 (-4.45)       \\
% & w/o     code review        &         75.00 (-4.27)      &    80.56 (-3.75)        \\ 
%                                     \bottomrule      
% \end{tabular}
% }
% \caption{Ablation study on code review and writing test suite}
% \label{tab:ablation_review_writing_tests}
% \end{table}

\paragraph{Impact of Code Dependency Graph} The Code Dependency Graph, $G$, plays a vital role in \methodnamews, as demonstrated by the results in Table \ref{tab:ablation-graph}. The variant of \methodname without the graph $G$ is susceptible to exceeding context length errors, while \methodname itself does not encounter this issue. Notably, the presence of $G$ leads to a substantial improvement in executability, increasing from 23.28\% to 57.50\%. The absence of the graph $G$ can result in a random order among testing scripts, causing inconsistencies during the bug-fixing process. Moreover, ignoring $G$ means that all source code is always included in the instructions, potentially overwhelming and even harming LLMs due to irrelevant information and increasing costs.

\begin{table}[ht]
\centering
\caption{Ablation study on the impact of $G$. \#ExceedingCL is the total number of Exceeding Context Length. In the case of the lack of $G$, we only consider tasks that do not encounter the Exceeding Context Length issue.}
\label{tab:ablation-graph}
\begin{tabular}{lcc}
\toprule
Statistical Index & \methodnamews & w/o $G$ \\ \midrule
Executability   & \textbf{57.50}  & 23.38   \\
Running Time (s)  & 465                          & \textbf{456}     \\
Token Usage       & \textbf{36818}                        & 37672   \\
Expenses (USD)    & \textbf{0.44}                         & 0.48    \\
\#Errors           & \textbf{0}                            & 10      \\
\#ExceedingCL     & \textbf{0}                            & 11     \\ \bottomrule
\end{tabular}

\end{table}

\section{Discussion \& Conclusion}
% In this paper, we introduced \methodnamews, a novel multi-agent software development framework inspired by Agile Methodology. By mimicking the Agile workflow and adapting it to a multi-agent context, \methodname enables dynamic adaptability and iterative enhancement capabilities. A key innovation of \methodname is the integration of static analysis through the Dynamic Code Graph Generator (DCGG), which dynamically produces a Code Dependency Graph (CDG) that captures the evolving relationships among code components.

In this paper, we introduce \methodnamews, a novel multi-agent software development framework inspired by Agile Methodology. Adapting Agile workflows to a multi-agent context, \methodname enhances dynamic adaptability and iterative development. A key innovation is the Dynamic Code Graph Generator (DCGG), which creates a Code Dependency Graph (CDG) to capture evolving code relationships.

Extensive evaluations on established benchmarks like HumanEval \citep{chen2021humaneval}, MBPP \citep{austin2021mbpp}, and our newly proposed benchmark, ProjectDev, confirm that \methodname achieves new state-of-the-art performance, outperforming recent SOTA models such as MetaGPT \citep{hong2024metagpt} and ChatDev \citep{qian2023chatdev}. The success of \methodname highlights the potential of integrating Agile Methodology and static analysis techniques into multi-agent software development frameworks.

% The inspiration drawn from professional human workflows in designing such agentic systems could pave the way for the future of multi-agent systems. Moreover, the concept of Sprint Planning in \methodname is similar to Dynamic Planning for agents, which aligns more closely with real-world scenarios compared to the single-plan approach commonly used in most agent systems.

% In conclusion, \methodname demonstrates the power of combining Agile Methodology, multi-agent systems, and static analysis to create a highly effective and adaptable software development framework. By drawing inspiration from professional human workflows and incorporating dynamic planning, \methodname represents a significant advancement in the field of multi-agent systems and software development automation.
Drawing from professional workflows, \methodname’s Sprint Planning mirrors Dynamic Planning for agents, making it more realistic than the single-plan approach typical in most systems. In conclusion, \methodname showcases the synergy of Agile Methodology, multi-agent systems, and static analysis, representing a significant advancement in software development automation.

\section*{Limitations}
While \methodname has demonstrated significant advancements in multi-agent software development, there are several areas for future work and limitations to be addressed.
One potential avenue for future research is the incorporation of additional Agile practices into the framework. For example, integrating pair programming or continuous integration and deployment (CI/CD) techniques could further enhance the collaboration and efficiency of the multi-agent system. Exploring the adaptation of other Agile methodologies, such as Kanban or Lean, could also provide valuable insights and improvements to the framework.

Another area for future work is the extension of \methodname to domains beyond software development. The principles of Agile Methodology and the multi-agent approach could be applied to other complex, iterative tasks such as product design, project management, or scientific research. Investigating the generalizability of the framework to these domains could lead to novel applications and advancements in various fields.

However, \methodname also has some limitations that should be acknowledged. One limitation is the reliance on LLMs for code generation and decision-making. While the integration of static analysis through the DCGG helps to mitigate some of the limitations of LLMs, there may still be cases where the generated code is suboptimal or fails to fully meet the requirements. Further research into improving the robustness and reliability of LLM-based code generation could help address this limitation.

Another limitation is the potential scalability issues when dealing with large, complex software projects. As the codebase grows, the computational resources required to maintain and update the CDG may become prohibitive. Future work could explore optimizations to the DCGG or alternative approaches to context retrieval that can handle larger-scale projects more efficiently.
Finally, the current implementation of \methodname focuses primarily on the technical aspects of software development, such as code generation and testing. However, real-world Agile development also involves important non-technical factors, such as team dynamics, communication, and project management. Incorporating these aspects into the multi-agent framework could provide a more comprehensive and realistic simulation of Agile software development.
Despite these limitations, \methodname represents a significant step forward in the automation of software development using multi-agent systems and Agile Methodology. By addressing these limitations and exploring the potential for future work, researchers can continue to push the boundaries of what is possible in this exciting field.

\bibliographystyle{plainnat}
\bibliography{bib/custom}

% \onecolumn
\appendix

\section{Appendix}
\subsection{Experimental Setting}\label{subsec:appendix-setting}
In our evaluation, we use GPT-3.5-Turbo-0613 and claude-3-haiku@20240307, and we set the temperature to 0.2, and $top\_p$ to 1, in accordance with previous work \citep{qian2023chatdev}. 

HumanEval \citep{chen2021humaneval} includes 164 handwritten programming problems, MBPP \citep{austin2021mbpp} is comprised of 427 Python tasks, and ProjectDev consists of 14 software development tasks.
For each dataset, every sample or task is executed three times, and the average results are reported, except for \#Errors and \#ExceedingCL, which are presented as total counts.
\subsection{More details on Analysis}
\subsubsection{Impact of The Three-step Code Review} \label{appendix:code-review}
In the Subsection \ref{subsec:implementation}, we introduce a three-step code review to conduct a static check for the correctness of source code. The detailed procedure is as below:
\begin{enumerate}[leftmargin=*]
	\item \textbf{Basic Implementation Checks:} Verification of implementation basics, such as the presence of empty methods (e.g., \textit{pass} in Python), adequacy of docstrings, and completeness of import statements.
	\item \textbf{Backlog Compliance:} Assessment of whether source code fulfills the tasks listed in the sprint backlog and includes all specified features.
	\item \textbf{Criteria Satisfaction and Bug Identification:} Final review to ensure the code meets the sprint's acceptance criteria and is free from potential bugs.
\end{enumerate}
We conduct an analysis to demonstrate the impact of our proposed three-step code review by comparing \methodname with its variant, where the three steps are condensed into a single step. Results in Table \ref{tab:ablation-code-review} show that breaking the code review into three steps improves accuracy. This is because the multi-step prompting strategy allows LLMs to perform static analyses more effectively and provide more precise feedback. In contrast, a single-step review involves instructions that encompass all desired constraints, making it more challenging for LLMs to understand and follow, thereby diminishing overall effectiveness.
\begin{table}[!ht]
\centering
\caption{Ablation study on our three-step prompting strategy for code review}
\label{tab:ablation-code-review}{
\begin{tabular}{llc}
\toprule
                  & {Model} & HumanEval            \\ \midrule
\multirow{2}{*}{GPT-3.5 Turbo} 
                                    & \methodnamews             &       70.53              \\ 
                                    & w/o    three-step review        &      67.68  (-2.85)            \\ \midrule
\multirow{2}{*}{Claude 3 Haiku}       
                                    & \methodnamews             &      79.27                 \\ 
                                     & w/o three-step review             &       75.61 (-3.66)
                                            \\ 
                                     \bottomrule      
\end{tabular}
}

\end{table}
% \subsubsection{Example of Code Dependency Graph}
\subsubsection{Error Bar}
In addition to main results in the Section \ref{subsec:results}, we also report error bars to demonstrate the robustness of our method, \methodnamews, under different runs. Results in Table \ref{tab:error-bar} show that
\methodname exhibits only minor variations across different runs.
\begin{table}[ht]
\centering
  \caption{Error bars of \methodname{} on HumanEval and MBPP}
    \label{tab:error-bar}
\begin{tabular}{lcc}
\toprule
\multirow{2}{*}{Model} & \multicolumn{2}{c}{Dataset} \\ \cmidrule(l){2-3} 

                                                         & HumanEval       & MBPP      \\ \midrule
{GPT-3.5 Turbo} 
                                               &       $70.53 \pm$    0.70    &     $80.92 \pm$ 0.83 \\ 
                                 \midrule
{Claude 3 Haiku}       
                                              &      $79.27 \pm$    0.86  &    $84.31 \pm$ 0.34    \\ 
                                    \bottomrule      
\end{tabular}

\end{table}
\subsubsection{Capabilities Analysis}
 In this subsection, we provide an analysis on capabilities of multi-agent frameworks for software engineering tasks. As presented in Table \ref{tab:comparison_agents}, compared to ChatDev \citep{qian2023chatdev} and MetaGPT \citep{hong2024metagpt}, \methodname has three additional capabilities, including flexible progress, incremental development, and modeling file dependencies. Incorporating these features can enhance the ability to perform software engineering tasks.
\begin{table}[!ht]
\centering
\caption{Comparison with other multi-agent frameworks}
\label{tab:comparison_agents}{
\begin{tabular}{lccc} 
\toprule
{Framework Capabiliy} & {ChatDev}             & {MetaGPT}             & {\methodnamews}           \\ 
\midrule
Code generation              & \textcolor{green}{\ding{51}} & \textcolor{green}{\ding{51}} & \textcolor{green}{\ding{51}}  \\
Role-based task management   & \textcolor{green}{\ding{51}} & \textcolor{green}{\ding{51}} & \textcolor{green}{\ding{51}}  \\
Code review                  & \textcolor{green}{\ding{51}} & \textcolor{green}{\ding{51}} & \textcolor{green}{\ding{51}}  \\
Flexible progress              & \textcolor{red}{\ding{55}}   & \textcolor{red}{\ding{55}}   & \textcolor{green}{\ding{51}}  \\
Incremental development     & \textcolor{red}{\ding{55}}   & \textcolor{red}{\ding{55}}   & \textcolor{green}{\ding{51}}  \\
Modeling file dependencies    & \textcolor{red}{\ding{55}}   & \textcolor{red}{\ding{55}}   & \textcolor{green}{\ding{51}}  \\
\bottomrule
\end{tabular}}

\end{table}
\subsubsection{Example of Code Dependency Graph} \label{appendix:example-graph}
\begin{figure}[ht]
    \centering
    \includegraphics[width=0.6\textwidth]{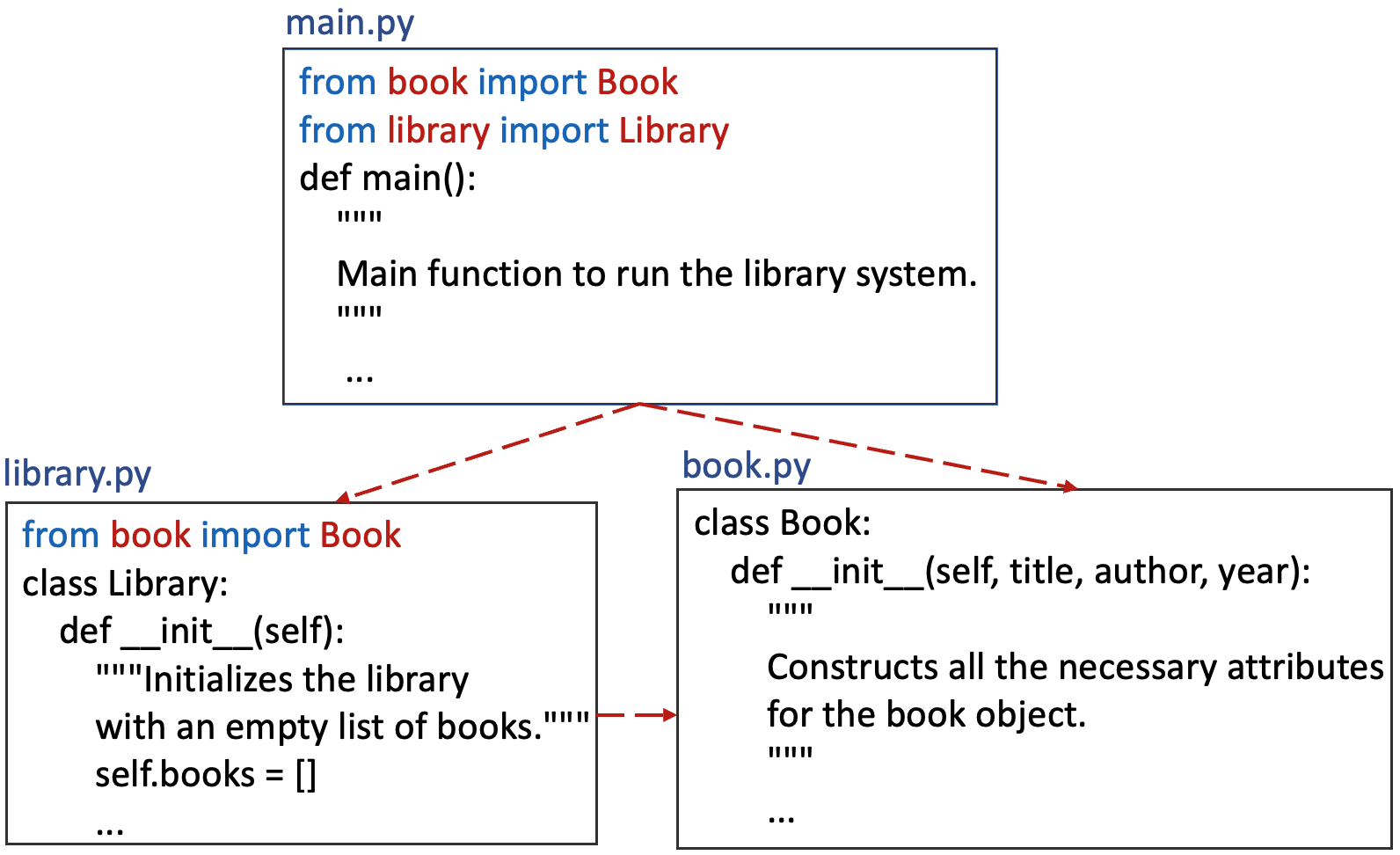}
    \caption{An example of the Code Dependency Graph}
    \label{fig:appendix-graph}
\end{figure}

In the Section \ref{sec:dcgg}, we propose a novel dependency graph $G$ modeling relationships among code files. Figure \ref{fig:appendix-graph} provides an illustrative example of the graph $G$ with each node representing a code file. The file \textit{library.py} is contingent on the file \textit{book.py}, so there is an edge from \textit{library.py} to  \textit{book.py}.

\subsection{Evaluation Details for ProjectDev} \label{appendix:evalution-projectdev}
In this section, we present our detailed evaluation process and the ProjectDev. We use GPT-3.5-Turbo-0613 as the backend model.
% %%%%%

%%%%
\subsubsection{Evaluation Steps} 
The entire evaluation process is conducted manually to ensure the correctness of the evaluation results. Participants are experienced developers with at least two years of Python programming experience. For each task, we run a method three times using the same prompt, producing a program for each run. For each generated program, we attempt to execute this program. If the program is executable, we evaluate it against all expected requirements. The final score is determined by the percentage of requirements the program meets. Below is the pseudocode for evaluating each task step-by-step:

\begin{algorithm}
\begin{algorithmic}[1]
\STATE \textbf{Input:} A task $t$ and its corresponding programs produced by ChatDev, MetaGPT, and \methodnamews
\STATE \textbf{Output:} Final score as a percentage

\STATE Load programs generated from ChatDev, MetaGPT, and \methodname that correspond to the task $t$

\STATE Load \textit{requirements} list of the task $t$ from the ProjectDev dataset
\STATE Initialize \textit{requirements\_met} to 0
\STATE Initialize \textit{total\_requirements} to length of \textit{requirements} list

\STATE Attempt to run each program
\IF {program is executable}
    \FOR {each requirement $r$ in \textit{requirements} list}
        \IF {the requirement $r$ is met}
            \STATE Increment \textit{requirements\_met} by 1
        \ENDIF
    \ENDFOR
\ENDIF

\STATE Calculate \textit{final\_score} as $(requirements\_met / total\_requirements) \times 100\%$

\STATE \textbf{return} \textit{final\_score}

\end{algorithmic}
\end{algorithm}

\subsubsection{ProjectDev dataset}
\textbf{Task Id: 1} \\
\textbf{Task: Snake game} \\
\textbf{Prompt: Create a snake game} \\ 
\textbf{Requirements:}
\begin{itemize}
  \item \textbf{Game Board:}
  \begin{itemize}[label={$\diamond$}]
    \item Create a grid-based game board.
    \item Define the dimensions of the grid (e.g., 10x10).
    \item Display the grid on the screen.
  \end{itemize}
  \item \textbf{Snake Initialization:}
  \begin{itemize}[label={$\diamond$}]
    \item Place the snake on the game board.
    \item Define the initial length and starting position of the snake.
    \item Choose a direction for the snake to start moving (e.g., right).
  \end{itemize}
  \item \textbf{Snake Movement:}
  \begin{itemize}[label={$\diamond$}]
    \item Implement arrow key controls for snake movement.
    \item Ensure the snake moves continuously in the chosen direction.
    \item Update the snake's position on the grid.
  \end{itemize}
  \item \textbf{Food Generation:}
  \begin{itemize}[label={$\diamond$}]
    \item Generate food at random positions on the game board.
    \item Ensure food doesn't appear on the snake's body.
  \end{itemize}
  \item \textbf{Collision Handling:}
  \begin{itemize}[label={$\diamond$}]
    \item Detect collisions between the snake and the game board boundaries.
    \item Detect collisions between the snake's head and its body.
    \item Detect collisions between the snake's head and the food.
  \end{itemize}
  \item \textbf{Snake Growth:}
  \begin{itemize}[label={$\diamond$}]
    \item Increase the length of the snake when it consumes food.
    \item Add a new segment to the snake's body.
  \end{itemize}
  \item \textbf{Score Display:}
  \begin{itemize}[label={$\diamond$}]
    \item Implement a scoring system.
    \item Display the current score on the screen.
  \end{itemize}
  \item \textbf{Game Over Condition:}
  \begin{itemize}[label={$\diamond$}]
    \item Trigger a game over scenario when the snake collides with the boundaries.
    \item Trigger a game over scenario when the snake collides with its own body.
    \item Display a game over message.
    \item Allow the player to restart the game.
  \end{itemize}
  \item \textbf{Graphics and User Interface:}
  \begin{itemize}[label={$\diamond$}]
    \item Use graphics or ASCII characters to represent the snake and food.
    \item Design a user-friendly interface with clear instructions and score display.
  \end{itemize}
  \item \textbf{Animations and Effects:}
  \begin{itemize}[label={$\diamond$}]
    \item Add animations for snake movement and growth.
    \item Implement visual effects for collisions and food consumption.
  \end{itemize}
\end{itemize}
\textbf{Task Id: 2} \\
\textbf{Task: Brick breaker game} \\
\textbf{Prompt: Create a brick breaker game} \\
\textbf{Requirements:}
\begin{itemize}
  \item \textbf{Game Board:}
  \begin{itemize}[label={$\diamond$}]
    \item Create a game board with a grid-based layout.
    \item Define the dimensions of the game board.
    \item Display the game board on the screen.
  \end{itemize}
  \item \textbf{Paddle Setup:}
  \begin{itemize}[label={$\diamond$}]
    \item Add a paddle at the bottom of the screen.
    \item Allow the player to control the paddle using keyboard or touch controls.
  \end{itemize}
  \item \textbf{Brick Formation:}
  \begin{itemize}[label={$\diamond$}]
    \item Generate a formation of bricks on the top of the screen.
    \item Define the number of rows and columns of bricks.
    \item Assign different colors or types to bricks.
  \end{itemize}
  \item \textbf{Ball Initialization:}
  \begin{itemize}[label={$\diamond$}]
    \item Place a ball on the paddle at the beginning of the game.
    \item Enable the player to launch the ball.
  \end{itemize}
  \item \textbf{Ball Movement:}
  \begin{itemize}[label={$\diamond$}]
    \item Implement physics for the ball's movement.
    \item Allow the ball to bounce off walls, paddle, and bricks.
    \item Update the ball's position continuously.
  \end{itemize}
  \item \textbf{Collision Detection:}
  \begin{itemize}[label={$\diamond$}]
    \item Detect collisions between the ball and the paddle.
    \item Detect collisions between the ball and the bricks.
    \item Handle different types of collisions appropriately.
  \end{itemize}
  \item \textbf{Brick Destruction:}
  \begin{itemize}[label={$\diamond$}]
    \item Remove a brick when the ball collides with it.
    \item Implement different points for different brick types.
    \item Track the number of bricks remaining.
  \end{itemize}
  \item \textbf{Scoring System:}
  \begin{itemize}[label={$\diamond$}]
    \item Implement a scoring system based on the bricks destroyed.
    \item Display the current score on the screen.
  \end{itemize}
  \item \textbf{Power-ups:}
  \begin{itemize}[label={$\diamond$}]
    \item Introduce power-ups that fall when certain bricks are destroyed.
    \item Implement power-ups such as extra balls, larger paddle, or slower ball speed.
  \end{itemize}
  \item \textbf{Game Levels:}
  \begin{itemize}[label={$\diamond$}]
    \item Create multiple levels with increasing difficulty.
    \item Design new brick formations for each level.
  \end{itemize}
  \item \textbf{Game Over Condition:}
  \begin{itemize}[label={$\diamond$}]
    \item End the game when the ball goes below the paddle.
    \item Display a game over message.
    \item Allow the player to restart the game.
  \end{itemize}
  \item \textbf{Graphics and User Interface:}
  \begin{itemize}[label={$\diamond$}]
    \item Use graphics to represent bricks, paddle, and ball.
    \item Design a user-friendly interface with clear instructions and score display.
  \end{itemize}
\end{itemize}
\textbf{Task Id: 3} \\
\textbf{Task: 2048 game} \\
\textbf{Prompt: Create a 2048 game} \\
\textbf{Requirements:}
\begin{itemize}
  \item \textbf{Game Board:}
  \begin{itemize}[label={$\diamond$}]
    \item Create a 4x4 grid as the game board.
    \item Display the grid on the web page.
  \end{itemize}
  \item \textbf{Tile Generation:}
  \begin{itemize}[label={$\diamond$}]
    \item Generate two initial tiles with the values 2 or 4 at random positions on the grid.
    \item Update the display to show the initial tiles.
  \end{itemize}
  \item \textbf{Tile Movement:}
  \begin{itemize}[label={$\diamond$}]
    \item Implement arrow key controls for tile movement (up, down, left, right).
    \item Allow tiles to slide in the chosen direction until they encounter the grid boundary or another tile.
    \item Combine tiles with the same value when they collide.
  \end{itemize}
  \item \textbf{Score Tracking:}
  \begin{itemize}[label={$\diamond$}]
    \item Implement a scoring system.
    \item Display the current score on the web page.
    \item Update the score when tiles are combined.
  \end{itemize}
  \item \textbf{Winning Condition:}
  \begin{itemize}[label={$\diamond$}]
    \item Define the winning condition as reaching the 2048 tile value.
    \item Display a victory message when the player reaches the winning condition.
    \item Allow the player to continue playing after winning.
  \end{itemize}
  \item \textbf{Game Over Condition:}
  \begin{itemize}[label={$\diamond$}]
    \item Implement a game over scenario when there are no more valid moves.
    \item Display a game over message.
    \item Allow the player to restart the game.
  \end{itemize}
  \item \textbf{Animations and Transitions:}
  \begin{itemize}[label={$\diamond$}]
    \item Add smooth animations for tile movements.
    \item Implement transitions for tile merging.
  \end{itemize}
  \item \textbf{Graphics and Styling:}
  \begin{itemize}[label={$\diamond$}]
    \item Use graphics or stylized numbers to represent different tile values.
  \end{itemize}
\end{itemize}
\textbf{Task Id: 4} \\
\textbf{Task: Flappy bird game} \\
\textbf{Prompt: Write code for Flappy Bird in python where you control a yellow bird continuously flying between a series of green pipes. The bird flaps every time you left click the mouse. If it falls to the ground or hits a pipe, you lose. This game goes on indefinitely until you lose; you get points the further you go.} \\
\textbf{Requirements:}
\begin{itemize}
  \item \textbf{Game Elements:}
  \begin{itemize}[label={$\diamond$}]
    \item Create a yellow bird as the main character.
    \item Design green pipes for the bird to navigate through.
    \item Set up a ground or background for the game.
  \end{itemize}
  \item \textbf{Bird Movement:}
  \begin{itemize}[label={$\diamond$}]
    \item Implement continuous forward movement of the bird.
    \item Make the bird fall due to gravity.
    \item Allow the bird to jump or "flap" when the left mouse button is clicked.
  \end{itemize}
  \item \textbf{Pipe Generation:}
  \begin{itemize}[label={$\diamond$}]
    \item Generate a series of pipes at regular intervals.
    \item Randomize the height of the gaps between pipes.
    \item Remove pipes from the screen when they move off the left side.
  \end{itemize}
  \item \textbf{Collision Detection:}
  \begin{itemize}[label={$\diamond$}]
    \item Detect collisions between the bird and the ground.
    \item Detect collisions between the bird and the pipes.
  \end{itemize}
  \item \textbf{Score Tracking:}
  \begin{itemize}[label={$\diamond$}]
    \item Implement a scoring system based on the distance traveled.
    \item Display the current score on the screen.
  \end{itemize}
  \item \textbf{Game Over Condition:}
  \begin{itemize}[label={$\diamond$}]
    \item Trigger a game over scenario when the bird hits the ground.
    \item Trigger a game over scenario when the bird hits a pipe.
    \item Display a game over message.
    \item Allow the player to restart the game.
  \end{itemize}
  \item \textbf{Animations and Effects:}
  \begin{itemize}[label={$\diamond$}]
    \item Add animations for bird flapping, pipe movement, and game over transitions.
    \item Implement visual effects for collisions.
  \end{itemize}
\end{itemize}
\textbf{Task Id: 5} \\
\textbf{Task: Tank battle game} \\
\textbf{Prompt: Create a tank battle game} \\
\textbf{Requirements:}
\begin{itemize}
  \item \textbf{Game Board:}
  \begin{itemize}[label={$\diamond$}]
    \item Create a grid-based game board representing the battlefield.
    \item Define the dimensions of the grid.
    \item Display the grid on the screen.
  \end{itemize}
  \item \textbf{Tank Initialization:}
  \begin{itemize}[label={$\diamond$}]
    \item Place two tanks on the game board for a two-player game.
    \item Set initial positions for each tank.
    \item Allow players to control their tanks using keyboard controls.
  \end{itemize}
  \item \textbf{Obstacles:}
  \begin{itemize}[label={$\diamond$}]
    \item Generate obstacles (e.g., walls, barriers) on the game board.
    \item Ensure obstacles are placed strategically to create a challenging battlefield.
  \end{itemize}
  \item \textbf{Tank Movement:}
  \begin{itemize}[label={$\diamond$}]
    \item Implement controls for tank movement (e.g., forward, backward, rotate left, rotate right).
    \item Allow tanks to move freely on the grid.
    \item Restrict tank movement when colliding with obstacles.
  \end{itemize}
  \item \textbf{Tank Firing:}
  \begin{itemize}[label={$\diamond$}]
    \item Implement controls for firing projectiles (e.g., bullets, missiles).
    \item Limit the firing rate to prevent spamming.
    \item Display projectiles on the screen.
  \end{itemize}
  \item \textbf{Projectile Collision:}
  \begin{itemize}[label={$\diamond$}]
    \item Detect collisions between projectiles and tanks.
    \item Deal damage to tanks when hit by projectiles.
    \item Remove projectiles when they collide with obstacles or go off the screen.
  \end{itemize}
  \item \textbf{Health and Damage:}
  \begin{itemize}[label={$\diamond$}]
    \item Assign health values to tanks.
    \item Display health bars for each tank.
    \item Trigger explosions or visual effects when a tank is destroyed.
  \end{itemize}
  \item \textbf{Scoring System:}
  \begin{itemize}[label={$\diamond$}]
    \item Implement a scoring system based on the number of tanks destroyed.
    \item Display the current score on the screen.
  \end{itemize}
  \item \textbf{Game Over Condition:}
  \begin{itemize}[label={$\diamond$}]
    \item Trigger a game over scenario when a tank's health reaches zero.
    \item Display a game over message.
    \item Allow players to restart the game.
  \end{itemize}
  \item \textbf{Graphics and User Interface:}
  \begin{itemize}[label={$\diamond$}]
    \item Use graphics to represent tanks, projectiles, and obstacles.
    \item Design a user-friendly interface with clear instructions and score display.
  \end{itemize}
  \item \textbf{Animations and Effects:}
  \begin{itemize}[label={$\diamond$}]
    \item Add animations for tank movement, firing, and explosions.
    \item Implement visual effects for collisions and explosions.
  \end{itemize}
\end{itemize}
\textbf{Task Id: 6} \\
\textbf{Task: Excel data process} \\
\textbf{Prompt: Write an excel data processing program based on streamlit and pandas. The screen first shows an excel file upload button. After the excel file is uploaded, use pandas to display its data content. The program is required to be concise, easy to maintain, and not over-designed. It uses streamlit to process web screen displays, and pandas is sufficient to process excel reading and display. Please make sure others can execute directly without introducing additional packages.} \\
\textbf{Requirements:}
\begin{itemize}
  \item \textbf{File Upload Button:}
  \begin{itemize}[label={$\diamond$}]
    \item Display a file upload button using any web library (such as Streamlit).
    \item Allow users to upload Excel files.
  \end{itemize}
  \item \textbf{Pandas Data Processing:}
  \begin{itemize}[label={$\diamond$}]
    \item Use Pandas to read the uploaded Excel file.
    \item Load the data into a Pandas DataFrame.
  \end{itemize}
  \item \textbf{Data Display:}
  \begin{itemize}[label={$\diamond$}]
    \item Display the content of the DataFrame.
    \item Show the first few rows of the data by default.
  \end{itemize}
  \item \textbf{Toggle for Full Data Display:}
  \begin{itemize}[label={$\diamond$}]
    \item Add a toggle button to switch between displaying the first few rows and the full data.
  \end{itemize}
  \item \textbf{Handling Missing Values:}
  \begin{itemize}[label={$\diamond$}]
    \item Check for and handle missing values in the data.
  \end{itemize}
  \item \textbf{Column Selection:}
  \begin{itemize}[label={$\diamond$}]
    \item Allow users to select specific columns for display.
    \item Display only the selected columns.
  \end{itemize}
  \item \textbf{Filtering:}
  \begin{itemize}[label={$\diamond$}]
    \item Implement simple data filtering based on user input.
    \item Display the filtered results.
  \end{itemize}
  \item \textbf{Sorting:}
  \begin{itemize}[label={$\diamond$}]
    \item Allow users to sort the data based on one or more columns.
    \item Display the sorted results.
  \end{itemize}
  \item \textbf{Download Processed Data:}
  \begin{itemize}[label={$\diamond$}]
    \item Provide a button to allow users to download the processed data.
  \end{itemize}
  \item \textbf{Visualizations (Optional):}
  \begin{itemize}[label={$\diamond$}]
    \item Include optional simple visualizations (e.g., bar chart, line chart).
  \end{itemize}
  \item \textbf{Error Handling:}
  \begin{itemize}[label={$\diamond$}]
    \item Implement error handling for file upload issues or data processing errors (e.g., file is not in excel or csv format).
    \item Display informative messages to the user.
  \end{itemize}
\end{itemize}
\textbf{Task Id: 7} \\
\textbf{Task: CRUD manage} \\
\textbf{Prompt: Write a management program based on the CRUD addition, deletion, modification and query processing of the customer business entity. The customer needs to save this information: name, birthday, age, sex, and phone. The data is stored in client.db, and there is a judgement whether the customer table exists. If it doesn’t, it needs to be created first. Querying is done by name; same for deleting. The program is required to be concise, easy to maintain, and not over-designed. The screen is realized through streamlit and sqlite—no need to introduce other additional packages.} \\
\textbf{Requirements:}
\begin{itemize}
  \item \textbf{Database Initialization:}
  \begin{itemize}[label={$\diamond$}]
    \item Connect to the SQLite database (client.db).
    \item Check if the customer table exists.
    \item If not, create the customer table with fields: name, birthday, age, sex, and phone.
  \end{itemize}
  \item \textbf{Add Customer (Create):}
  \begin{itemize}[label={$\diamond$}]
    \item Provide input fields for name, birthday, age, sex, and phone.
    \item Allow users to add a new customer to the database.
    \item Validate input data (e.g., check if the phone is valid).
  \end{itemize}
  \item \textbf{Query Customer (Read):}
  \begin{itemize}[label={$\diamond$}]
    \item Implement a query interface with an input field for the customer's name.
    \item Display the customer information if found.
    \item Provide a message if the customer is not found.
  \end{itemize}
  \item \textbf{Update Customer (Modify):}
  \begin{itemize}[label={$\diamond$}]
    \item Allow users to update customer information.
    \item Display the current information and provide input fields for modifications.
    \item Update the database with the modified data.
  \end{itemize}
  \item \textbf{Delete Customer (Delete):}
  \begin{itemize}[label={$\diamond$}]
    \item Allow users to delete a customer based on their name.
    \item Display a confirmation message before deleting.
    \item Update the database by removing the customer.
  \end{itemize}
  \item \textbf{Display All Customers:}
  \begin{itemize}[label={$\diamond$}]
    \item Create a section to display all customers in the database.
    \item Display relevant information for each customer.
  \end{itemize}
  \item \textbf{UI:}
  \begin{itemize}[label={$\diamond$}]
    \item Design a (Streamlit) user interface with a clean layout.
    \item Use (Streamlit) components for input fields, buttons, and information display.
  \end{itemize}
  \item \textbf{Error Handling:}
  \begin{itemize}[label={$\diamond$}]
    \item Implement error handling for database connectivity issues.
    \item Provide user-friendly error messages.
  \end{itemize}
\end{itemize}
\textbf{Task Id: 8} \\
\textbf{Task: Custom press releases} \\
\textbf{Prompt: Create custom press releases; develop a Python script that extracts relevant information about company news from external sources, such as social media; extract update interval database for recent changes. The program should create press releases with customizable options and export writings to PDFs, NYTimes API JSONs, media format styled with interlink internal fixed character-length metadata.} \\
\textbf{Requirements:}
\begin{itemize}
  \item \textbf{Data Extraction from External Sources:}
  \begin{itemize}[label={$\diamond$}]
    \item Implement web scraping or use APIs to extract relevant information from social media and other external sources.
    \item Extract data such as company updates, news, and events.
  \end{itemize}
  \item \textbf{Update Interval Database:}
  \begin{itemize}[label={$\diamond$}]
    \item Develop a database to store information about recent changes or updates.
    \item Include fields like timestamp, source, and content.
  \end{itemize}
  \item \textbf{Customizable Press Release Options:}
  \begin{itemize}[label={$\diamond$}]
    \item Design a user interface or command-line options for users to customize press releases.
    \item Allow customization of content, format, and metadata.
  \end{itemize}
  \item \textbf{Press Release Content Generation:}
  \begin{itemize}[label={$\diamond$}]
    \item Develop algorithms to generate coherent and concise press release content.
    \item Use extracted information to create engaging narratives.
  \end{itemize}
  \item \textbf{Export Options:}
  \begin{itemize}[label={$\diamond$}]
    \item Provide options to export press releases in different formats, such as PDF and NYTimes API JSONs.
    \item Include customizable templates for different media formats.
  \end{itemize}
  \item \textbf{Metadata Inclusion:}
  \begin{itemize}[label={$\diamond$}]
    \item Add metadata to the press releases, including fixed character-length metadata.
    \item Ensure metadata includes relevant information such as publication date, source, and author.
  \end{itemize}
  \item \textbf{PDF Export:}
  \begin{itemize}[label={$\diamond$}]
    \item Implement functionality to export press releases as PDF files.
    \item Allow users to specify PDF export options (e.g., layout, fonts).
  \end{itemize}
  \item \textbf{NYTimes API JSON Export:}
  \begin{itemize}[label={$\diamond$}]
    \item Integrate with the NYTimes API to fetch additional relevant information.
    \item Format and export the press releases in JSON format compatible with the NYTimes API.
  \end{itemize}
  \item \textbf{Media Format Styling:}
  \begin{itemize}[label={$\diamond$}]
    \item Apply styling to the press releases based on different media formats.
    \item Ensure the styling is consistent with industry standards.
  \end{itemize}
  \item \textbf{Interlink Internal Content:}
  \begin{itemize}[label={$\diamond$}]
    \item Implement interlinking of internal content within the press releases.
    \item Include links to relevant articles, documents, or resources.
  \end{itemize}
  \item \textbf{Error Handling:}
  \begin{itemize}[label={$\diamond$}]
    \item Implement error handling mechanisms for data extraction, customization, and export processes.
    \item Provide clear error messages and logging.
  \end{itemize}
\end{itemize}
\textbf{Task Id: 9} \\
\textbf{Task: Caro game} \\
\textbf{Prompt: Create a caro game in python} \\
\textbf{Requirements:}
\begin{itemize}
  \item \textbf{Game Board:}
  \begin{itemize}[label={$\diamond$}]
    \item Create a grid-based game board for Caro.
    \item Define the dimensions of the board (commonly 15x15 for Caro).
    \item Display the game board on the console or a graphical user interface.
  \end{itemize}
  \item \textbf{Player vs. Player:}
  \begin{itemize}[label={$\diamond$}]
    \item Implement a two-player mode where two human players can take turns.
    \item Allow players to place their marks (X or O) on the board.
  \end{itemize}
  \item \textbf{Winning Conditions:}
  \begin{itemize}[label={$\diamond$}]
    \item Detect and announce a winner when a player gets five marks in a row horizontally, vertically, or diagonally.
    \item Declare a draw when the board is full and no player has won.
  \end{itemize}
  \item \textbf{User Interface:}
  \begin{itemize}[label={$\diamond$}]
    \item Create a user-friendly interface for players to interact with the game.
    \item Display the current state of the board after each move.
  \end{itemize}
  \item \textbf{Input Handling:}
  \begin{itemize}[label={$\diamond$}]
    \item Implement input handling for player moves.
    \item Ensure valid moves and handle invalid inputs gracefully.
  \end{itemize}
  \item \textbf{Restart and Exit Options:}
  \begin{itemize}[label={$\diamond$}]
    \item Provide options to restart the game or exit the program after a game is completed.
    \item Ask for confirmation before restarting or exiting.
  \end{itemize}
  \item \textbf{Game Logic:}
  \begin{itemize}[label={$\diamond$}]
    \item Implement the core game logic, including checking for winning conditions, updating the board, and managing turns.
  \end{itemize}
\end{itemize}
\textbf{Task Id: 10} \\
\textbf{Task: Video player} \\
\textbf{Prompt: Create a video player in python that can play videos from users. The application can also support multiple controls such as play, pause and stop.} \\
\textbf{Requirements:}
\begin{itemize}
  \item \textbf{User Interface:}
  \begin{itemize}[label={$\diamond$}]
    \item Develop a user interface to display the video player.
    \item Include a section to display the video content.
    \item Create a space for control buttons (play, pause, stop).
  \end{itemize}
  \item \textbf{Video Loading:}
  \begin{itemize}[label={$\diamond$}]
    \item Implement functionality to load videos from user input.
    \item Support common video formats (e.g., MP4, AVI).
    \item Handle errors gracefully if the video format is not supported.
  \end{itemize}
  \item \textbf{Play Button:}
  \begin{itemize}[label={$\diamond$}]
    \item Implement a "Play" button to start playing the loaded video.
    \item Ensure the button is responsive and updates its state (e.g., changes to a pause button when the video is playing).
  \end{itemize}
  \item \textbf{Pause Button:}
  \begin{itemize}[label={$\diamond$}]
    \item Implement a "Pause" button to temporarily pause the video.
    \item Allow users to resume playback from the paused state.
  \end{itemize}
  \item \textbf{Stop Button:}
  \begin{itemize}[label={$\diamond$}]
    \item Implement a "Stop" button to stop the video playback.
    \item Reset the video to the beginning when stopped.
  \end{itemize}
  \item \textbf{Time Slider:}
  \begin{itemize}[label={$\diamond$}]
    \item Add a time slider or progress bar to visualize the current position in the video.
    \item Allow users to click on the slider to jump to specific points in the video.
  \end{itemize}
  \item \textbf{Error Handling:}
  \begin{itemize}[label={$\diamond$}]
    \item Implement error handling for cases where the video fails to load or encounters playback issues.
    \item Display informative error messages to users.
  \end{itemize}
\end{itemize}
\textbf{Task Id: 11} \\
\textbf{Task: Youtube Video Downloader} \\
\textbf{Prompt: Create a Youtube Video Downloader in Python that receives a Youtube link as an input and then downloads the video with multiple options of resolutions} \\
\textbf{Requirements:}
\begin{itemize}
  \item \textbf{User Input:}
  \begin{itemize}[label={$\diamond$}]
    \item Develop a user interface or command-line interface to accept a YouTube video link as input.
  \end{itemize}
  \item \textbf{YouTube API Integration (Optional):}
  \begin{itemize}[label={$\diamond$}]
    \item Integrate with the YouTube API to fetch information about the available video resolutions and formats.
    \item Retrieve details such as video title, available resolutions, and formats.
  \end{itemize}
  \item \textbf{Video Resolution Options:}
  \begin{itemize}[label={$\diamond$}]
    \item Present the user with options to choose from various video resolutions.
    \item Display information about each resolution (e.g., resolution, format, file size).
  \end{itemize}
  \item \textbf{Download Mechanism:}
  \begin{itemize}[label={$\diamond$}]
    \item Implement the video download mechanism using a library like pytube or similar.
    \item Allow users to choose the desired video resolution.
  \end{itemize}
  \item \textbf{Download Progress Display:}
  \begin{itemize}[label={$\diamond$}]
    \item Display a progress bar or percentage to show the download progress.
    \item Update the progress in real-time during the download.
  \end{itemize}
  \item \textbf{File Naming Options:}
  \begin{itemize}[label={$\diamond$}]
    \item Provide options for users to specify the name of the downloaded file.
    \item Generate a default name based on the video title.
  \end{itemize}
  \item \textbf{Download Location:}
  \begin{itemize}[label={$\diamond$}]
    \item Allow users to specify the directory where the video will be saved.
    \item Use a default directory if the user doesn't specify one.
  \end{itemize}
  \item \textbf{Video Information Display:}
  \begin{itemize}[label={$\diamond$}]
    \item Display relevant information about the video, such as title, duration, and uploader.
    \item Show information before and after the download.
  \end{itemize}
  \item \textbf{Error Handling:}
  \begin{itemize}[label={$\diamond$}]
    \item Implement error handling to gracefully handle issues such as invalid URLs, network errors, or download failures.
  \end{itemize}
\end{itemize}
\textbf{Task Id: 12} \\
\textbf{Task: QR Code Generator and Detector} \\
\textbf{Prompt: Create a Python program that produces a QR code for the input from users and decodes a QR code from users.} \\
\textbf{Requirements:}
\begin{itemize}
  \item \textbf{Generating QR Codes:}
  \begin{itemize}[label={$\diamond$}]
    \item \textbf{User Input:}
    \begin{itemize}[label={$\diamond$}]
      \item Develop a user interface or command-line interface to accept user input.
      \item Allow users to input text or a URL for which a QR code will be generated.
    \end{itemize}
    \item \textbf{QR Code Generation Library:}
    \begin{itemize}[label={$\diamond$}]
      \item Choose a QR code generation library in Python (e.g., qrcode).
    \end{itemize}
    \item \textbf{QR Code Generation:}
    \begin{itemize}[label={$\diamond$}]
      \item Generate QR code successfully with valid input.
      \item Allow users to customize QR code parameters (e.g., size, color).
    \end{itemize}
    \item \textbf{Display QR Code:}
    \begin{itemize}[label={$\diamond$}]
      \item Display the generated QR code to the user on screen.
    \end{itemize}
    \item \textbf{Save QR Code:}
    \begin{itemize}[label={$\diamond$}]
      \item Implement an option for users to save the generated QR code as an image file (e.g., PNG).
    \end{itemize}
  \end{itemize}
  \item \textbf{Decoding QR Codes:}
  \begin{itemize}[label={$\diamond$}]
    \item \textbf{User Input for Decoding:}
    \begin{itemize}[label={$\diamond$}]
      \item Allow users to input an image file containing a QR code for decoding.
      \item Support common image formats (e.g., PNG, JPEG).
    \end{itemize}
    \item \textbf{QR Code Decoding Library:}
    \begin{itemize}[label={$\diamond$}]
      \item Choose a QR code decoding library in Python (e.g., opencv, pyzbar).
    \end{itemize}
    \item \textbf{QR Code Decoding:}
    \begin{itemize}[label={$\diamond$}]
      \item Decode QR code successfully with valid input.
      \item Display the decoded text or URL to the user.
    \end{itemize}
    \item \textbf{Error Handling:}
    \begin{itemize}[label={$\diamond$}]
      \item Implement error handling for cases where decoding fails or the input is invalid.
    \end{itemize}
  \end{itemize}
\end{itemize}
\textbf{Task Id: 13} \\
\textbf{Task: To-Do List App} \\
\textbf{Prompt: Create a simple to-do list application in python where users can add, edit, and delete tasks. The application includes some features like marking tasks as completed and categorizing tasks.} \\
\textbf{Requirements:}
\begin{itemize}
  \item \textbf{User Interface:}
  \begin{itemize}[label={$\diamond$}]
    \item Provide options for adding, editing, deleting, marking as completed, and categorizing tasks.
  \end{itemize}
  \item \textbf{Task List Display:}
  \begin{itemize}[label={$\diamond$}]
    \item Display the list of tasks in a readable format.
    \item Include information such as task name, status (completed or not), and category.
  \end{itemize}
  \item \textbf{Adding Tasks:}
  \begin{itemize}[label={$\diamond$}]
    \item Implement functionality to add new tasks to the to-do list.
    \item Allow users to input the task name, status, and category.
  \end{itemize}
  \item \textbf{Editing Tasks:}
  \begin{itemize}[label={$\diamond$}]
    \item Provide an interface to modify task details such as name, status, and category.
  \end{itemize}
  \item \textbf{Deleting Tasks:}
  \begin{itemize}[label={$\diamond$}]
    \item Implement the ability to delete tasks from the to-do list.
    \item Confirm user intent before deleting a task.
  \end{itemize}
  \item \textbf{Marking Tasks as Completed:}
  \begin{itemize}[label={$\diamond$}]
    \item Allow users to mark tasks as completed.
    \item Toggle their completion status.
  \end{itemize}
  \item \textbf{Categorizing Tasks:}
  \begin{itemize}[label={$\diamond$}]
    \item Filter tasks by category.
  \end{itemize}
  \item \textbf{Saving and Loading Tasks:}
  \begin{itemize}[label={$\diamond$}]
    \item Save the to-do list data to a file (e.g., JSON or CSV).
    \item Load the saved data when the application starts.
  \end{itemize}
\end{itemize}
\textbf{Task Id: 14} \\
\textbf{Task: Calculator} \\
\textbf{Prompt: Create a calculator with python that performs basic arithmetic operations. Practices user input, functions, and mathematical operations.} \\
\textbf{Requirements:}
\begin{itemize}
  \item \textbf{User Interface:}
  \begin{itemize}[label={$\diamond$}]
    \item Create an interface for basic arithmetic operations (addition, subtraction, multiplication, division).
  \end{itemize}
  \item \textbf{User Input:}
  \begin{itemize}[label={$\diamond$}]
    \item Accept user input for numeric values and mathematical operations.
    \item Allow multiple input numbers and operations.
  \end{itemize}
  \item \textbf{Arithmetic Operations:}
  \begin{itemize}[label={$\diamond$}]
    \item Successfully implement functions for addition operations.
    \item Successfully implement functions for subtraction operations.
    \item Successfully implement functions for multiplication operations.
    \item Successfully implement functions for division operations.
  \end{itemize}
  \item \textbf{Functionality:}
  \begin{itemize}[label={$\diamond$}]
    \item Allow users to choose the operation they want to perform.
  \end{itemize}
  \item \textbf{Calculation Execution:}
  \begin{itemize}[label={$\diamond$}]
    \item Display the result of the calculation.
  \end{itemize}
  \item \textbf{Continuous Calculation (Optional):}
  \begin{itemize}[label={$\diamond$}]
    \item Optionally, allow users to perform continuous calculations without restarting the program.
  \end{itemize}
  \item \textbf{Error Handling:}
  \begin{itemize}[label={$\diamond$}]
    \item Implement error handling for cases where the user provides invalid input or attempts an unsupported operation.
  \end{itemize}
\end{itemize}

\end{document}